\documentclass[
 reprint,
 amsmath,amssymb,
prb
]{revtex4-2}
\usepackage[export]{adjustbox} 
\usepackage{graphicx}
\usepackage{dcolumn}
\usepackage{bm}

\usepackage{graphicx}
\graphicspath{ {./Figures/} }
\usepackage{xfrac}
\usepackage{amsmath}
\usepackage{amssymb}
\usepackage{amsfonts}
\usepackage{caption}
\usepackage{titlesec}
\titlespacing{\section}{0pt}{\baselineskip}{\baselineskip}
\usepackage{diagbox} 
\usepackage{subcaption}
\usepackage{float}
\usepackage{color}
\usepackage{mathtools}
\usepackage{hyperref}
\usepackage{url}
\usepackage{bm}
\usepackage{eqnarray}

\newcommand*\dif{\mathop{}\!\mathrm{d}}

\usepackage{amsthm}

\usepackage[table,xcdraw]{xcolor}

\usepackage[makeroom]{cancel} 
\hypersetup{
  colorlinks   = true, 
  urlcolor     = blue, 
  linkcolor    = blue, 
  citecolor    = red 
}
\usepackage{braket}
\usepackage[table,xcdraw]{xcolor}
\usepackage{color, colortbl}
\usepackage{relsize}

\begin{document}

\setlength{\abovedisplayskip}{5pt}
\setlength{\belowdisplayskip}{5pt}

\preprint{APS/123-QED}

\title{A Quantized Interband Topological Index in Two-Dimensional Systems}

\author{Tharindu Fernando}
 \email{tharindu@uw.edu}
 \affiliation{
 Department of Physics, University of Washington, Seattle, WA 98195 USA
}
\author{Ting Cao}
 \affiliation{Department of Materials Science and Engineering, University of Washington, Seattle, WA 98195 USA}

\date{\today}

\begin{abstract}
We introduce a novel gauge-invariant, quantized interband index in two-dimensional (2D) multiband systems. It provides a bulk topological classification of a submanifold of parameter space (e.g., an electron valley in a Brillouin zone), and therefore overcomes difficulties in characterizing topology of submanifolds. We confirm its topological nature by numerically demonstrating a one-to-one correspondence to the valley Chern number in $k\cdot p$ models (e.g., gapped Dirac fermion model), and the first Chern number in lattice models (e.g., Haldane model). Furthermore, we derive a band-resolved topological charge and demonstrate that it can be used to investigate the nature of edge states due to band inversion in valley systems like multilayer graphene. 
\end{abstract}

         
\maketitle

\newpage

Topological and geometric effects are being heavily investigated
in contemporary condensed matter physics. 
For an adiabatic evolution along a closed loop in a 2D parameter space, 
the geometric part of the final electronic eigenstate's phase is $U(1)$ gauge-invariant
(modulo $2\pi$). This Berry phase contribution depends solely on the 
geometry of the parameter space \cite{berry1984quantal}. The corresponding Berry curvature is
a geometrically local quantity, which when summed over the entire 2D-space manifold,
may yield topological quantities such as the first Chern number
\cite{thouless1982quantized,simon1983holonomy,nakahara2018geometry}.
In solid state physics, the Berry phase also plays vital roles in
topology-related phenomena, and applications including
electric polarization, orbital magnetism, 
adiabatic charge pumping, 
various types of Hall effects, and edge state engineering \cite{xiao2010berry,hasan2010colloquium}.

Despite these advancements, the understanding of the multi-level topology of parameter-space submanifolds is arguably still under development. 
This Article will focus on $k$-space submanifolds in the vicinity of band edges at high-symmetry points (or so-called \emph{valleys}) \cite{fan2022tracking}. These valley degrees of freedom play key roles in future electronics and quantum information science, as quasiparticles residing in the valleys may carry information much like charge and spin \cite{rycerz2007valley,jung2011valley,zhang2011spontaneous,xiao2007valley,neto2009electronic,gorbachev2014detecting,xu2014spin,mak2014valley,semenoff2008domain,martin2008topological,yao2009edge,li2010marginality,li2011topological,qiao2011electronic,zhang2013valley,vaezi2013topological,ju2015topological,li2016gate,fan2022tracking,qiao2013topological,ezawa2014symmetry}. 
The associated topology is currently studied using the \emph{valley Chern number}. 
This is usually calculated using a loop integral of the Berry connection (a method that is arguably restrictive due to requiring a non-singular gauge), or by integrating Berry curvature, in the vicinity of a valley
\cite{xiao2010berry}. 
Generally, both $k\cdot p$ and lattice models have proven useful in the study of topological phenomena of valleys. 
However, in $k\cdot p$ models, the area of Berry curvature integration required to obtain quantized valley Chern number is infinite (or equivalently, requires an infinitesimally small band gap \cite{fan2022tracking,zhang2013valley,xiao2010berry}). 
On the other hand, in lattice models, there is no general quantized character to describe valley topology when the Berry curvature is not peaked at the valley; at least not without low-energy expansions or additional synthetic dimensions \cite{hasan2010colloquium,fradkin2013field}. 
In addition, relating existing bulk indices to edge modes by the bulk-edge correspondence often requires summing the valley Chern number over all filled bands \cite{xiao2010berry}, and/or downfolding multiband Hamiltonians into simpler models \cite{lowdin1951note, luttinger1955motion, bir1974symmetry}. 
This may cause a loss of information on the topological origin of edge states. 
For example, if the edge state arises from inverting a pair of bands among many bands, such band-resolved information would be missing in the valley Chern number description.

In this Article, we introduce a new topological index $\Theta$, and the \emph{interband frequency}; a correction term that keeps $\Theta$ quantized. Our approach gives us a \emph{meaningful} topological valley index using a \emph{finite} $k$-space integration, in \emph{both} $k\cdot p$ and lattice models. We also present a band-resolved topological charge $\Xi$, that identifies orbitals associated with band inversions \emph{without} downfolding multiband Hamiltonians.

\emph{Interband index in 2D.}---Consider the time-independent Schr\"{o}dinger equation for 
an $N$-level non-degenerate Hamiltonian $H(\boldsymbol{k})$ over 2D parameter space $k$:
$
    {H}(\boldsymbol{k})\ket{m(\boldsymbol{k})} = E_m (\boldsymbol{k}) \ket{m(\boldsymbol{k})},
    (m = 1,2, ... , N),
$
where $\ket{m}$ are orthonormal instantaneous eigenstates with eigenvalues $E_m (\boldsymbol{k})$.
For an adiabatic evolution along a closed $k$-space loop $\partial{ \mathcal{M}}$,
we define the interband index $\Theta_{\boldsymbol{k}}$, following the definition of the \textit{interlevel character} in Ref.  \cite{xu2018quantized}: 
\begin{equation}
	\label{eq:theta:k}
	2\pi\Theta_{mn} = \Delta\Phi_{mn}
	- \oint\limits_{\textrm{ }\partial{ \mathcal{M}}}
	\dif \arg\Braket{{m} | \boldsymbol{\nabla_k} n}\cdot
	\hat{e}_{\tau}, 
\end{equation}

\begin{figure}[h]
  \centering
  \includegraphics[width=0.425\textwidth]{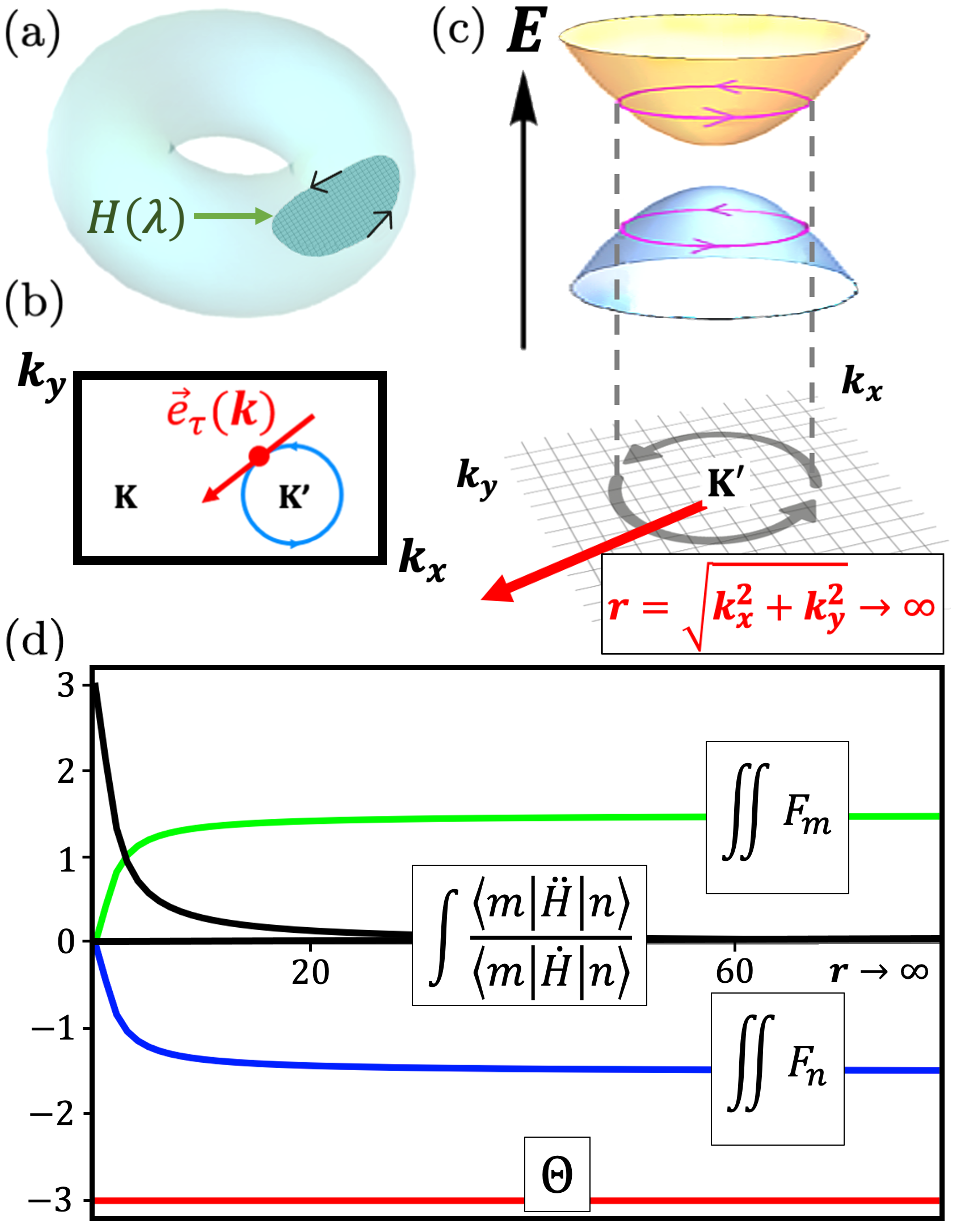}
  \caption{    
        {(a)} 2D $k$-space Brillouin torus 
        for a two-level system. 
        The counterclockwise closed loop is $\partial\mathcal{M}$,
        and defines
        the shaded region of the torus
        as $\mathcal{M}$, by convention. 
        For a $k$-space loop parameterized by $\lambda$,
        the adiabatic evolution is given by 
        ${H}(\lambda)  \equiv{H}(\boldsymbol{k})$.
        {(b)} For $k$ constrained to 
        $\partial\mathcal{M}$,
        the tangential vector $\vec{e}_\tau (k)$
        at a point
        is denoted in red.
        K and K' are high-symmetry points. 
        {(c)} Two energy levels of the dispersion \textbf{\emph{E}}
        in the vicinity of the K' valley. 
        The red arrow schematically illustrates how the gray $k$-space loop's radius $r$ is increased in (d).
        {(d)} Berry curvature area integrals of upper and lower bands, integral of interband frequency, and the interband index in Eq. \eqref{eq:freq:onlyF}
    as we vary the size of $k$ loops using $r$. 
    }
    \label{fig:intro:interband}
\end{figure}

Above, we used the definitions: 
 $\boldsymbol{k}=(k_x,k_y)$; $\dif$ is the total derivative with respect to $k_x$ and $k_y$;
$\boldsymbol{\nabla_k}=(\partial_{k_x},\partial_{k_y})$;
$\hat{e}_{\tau}={\boldsymbol{\dot{k}}}/{|\boldsymbol{\dot{k}|}}$ 
is the unit tangential operator at a point
on the loop $\partial{ \mathcal{M}}$ (see Fig. \ref{fig:intro:interband} (b));
$\boldsymbol{\dot{k}}={\dif{\boldsymbol{k}(\lambda)}}/{\dif{\lambda}}$ for some $\lambda$ that
parameterizes the loop $\boldsymbol{k}=(k_x(\lambda),k_y(\lambda))$;
and $\Delta\Phi_{mn}=\Phi_{m} - \Phi_{n}$,
where $\Phi_m=\int_{\partial{\mathcal{M}}}
	 \mathcal{A}_{m}^\mu \dif\lambda_{\mu}
	 -
	 \iint_{{\mathcal{M}}}
	 F_m\dif \lambda_{\mu}\dif\lambda_{\nu}.$
For brevity, we henceforth drop the 
differential elements $\dif\lambda_{\mu}$.
$\Phi_m$ is the number of
\emph{Berry singularities} in level $|m\rangle$:
It
is the difference between the 
the line integral of the standard 
Berry connection
$\mathcal{A}_{m}^\mu = i \braket{{m} | \frac{\partial}{\partial \lambda_{\mu}} m}$
along $\partial{ \mathcal{M}}$,
and the area integral of the 
Berry curvature $F_m = \frac{\partial}{\partial \lambda_{\mu}} \mathcal{A}_{m}^\nu 
	 - 
\frac{\partial}{\partial \lambda_{\nu}} \mathcal{A}_{m}^\mu$
over the region $\mathcal{M}$ specified by the loop.
So, $\Phi_m$ could be interpreted as the
quantized `amount' by which
Stokes' theorem fails.
Then, $\Delta\Phi$ is the net number of Berry singularities
between the levels considered. 
Notice
that in the case without gauge singularities,
$\Phi_m $ reduces to $0$ as $\int_{\partial{\mathcal{M}}}\mathcal{A}_{m}^\mu =\iint_{{\mathcal{M}}}F_m$.

Following the derivation in Supplementary Material (SM) \ref{app:expanddarg},
$\Theta_{mn}$ may be written as:
\begin{align}
\begin{split}
\label{eq:freq:onlyF}
2&\pi\Theta_{mn} = 
{\iint\limits_{{\mathcal{M}}}
	 F_{n} 
	 - \iint\limits_{{\mathcal{M}}}
	 F_m-
	 \oint\limits_{{\partial\mathcal{M}}} 
\text{Im}\frac{\langle m |\ddot{H} | n\rangle}{\langle m|\dot{H}|n\rangle}}\\
&-
\oint\limits_{{\partial\mathcal{M}}}
\sum\limits_{q\neq m,n}\frac{1}{E_{nq}}\left(2 - \frac{E_{nm}}{E_{qm}}\right)\text{Im}\frac{\langle m|\dot{H}|q\rangle\langle q|\dot{H}|n\rangle}{\langle m |\dot{H}|n\rangle}.
\end{split}
\end{align}
The overhead dots represent derivation with respect to parameter $\lambda$.
All terms in Eq. 
\eqref{eq:freq:onlyF} are gauge-independent,
and so, potentially observable.
The first two terms in the right hand side of Eq. \eqref{eq:freq:onlyF}
make the difference
between the Berry curvature integrals.
The third boundary term includes 
$\langle m |\ddot{H} | n\rangle / {\langle m|\dot{H}|n\rangle}$, which we call the \emph{interband frequency},
since it 
resembles the ratio of an acceleration-like 
quantity to a velocity-like quantity.
Since $k\rightarrow\infty$ implies the loop-parameter (e.g., time) $\lambda\rightarrow\infty$, we intuit that the frequency ($\propto 1/\lambda$) 
$\rightarrow 0$. We show later in Fig. \ref{fig:intro:interband} (d) that as $k\rightarrow\infty$, this 
correction term also tends to $0$ in our numerical calculations on 2-band models.
To our knowledge, the interband 
frequency is new to the literature.
Due to its dependence on the tangential vector
$\hat{e}_{\tau}$, 
a unique vector field 
 may be defined only \emph{after} a loop is chosen.
This makes 
the individual terms in the interband frequency ratio 
differ from existing quantities in the literature (such as the interband acceleration in second order nonlinear responses \cite{holder2020consequences,kitamura2020nonreciprocal}).

The physical significance of the interband index thus becomes clear: It is a quantized topological character for a submanifold of 2D parameter space
that depicts the difference between the Berry phases of a pair of bands, corrected by the interband frequency and other terms.
Next, we demonstrate the physical meaning and applications of the interband index 
in $k\cdot p$ and lattice models.

\emph{Application to the
gapped Dirac fermion model and Haldane model.}---We first calculate $\Theta$ and compare it with 
existing topological characterizations of 
$k\cdot p$ models. While we use the gapped Dirac fermion model for illustration, our results hold for the other systems we tested (SM \ref{app:topo_quantities}).
Effective models often follow from
low-energy expansions about a high-symmetry point or band extremum $P$,
and can describe important band inversions leading to chiral edge states \cite{yao2009edge,ju2015topological,zhang2011spontaneous,neto2009electronic,jung2011valley,martin2008topological,zhang2013valley,vaezi2013topological,hasan2010colloquium}. 
However, being subspaces of the complete 
Hilbert space, these models 
may not have a closed 
$k$-space manifold (e.g., 2D Brillouin torus).
One significant example is the electron valley degree of freedom.
The conventional valley Chern number $\bar{\nu}^P$ 
is usually given by the $k$-space integral 
of the Berry curvature $F_m$
of filled bands in the vicinity of a valley centered at $P$, integrated to infinity:
\begin{equation}
    \label{eq:ValleyC_Infinite}
    \bar{\nu}^P = \mathlarger{\sum}_{i\in\text{filled}}\quad\iint\limits_{k\rightarrow\infty} F_i \equiv \mathlarger{\sum}_{i\in\text{filled}}\bar{C}_i^{P}.
\end{equation}
Above, $\bar{C}_i^{P}$ is the valley Chern number at $P$ per band $\ket{i}$, and 
the overhead bar indicates that we used the conventional definition of \eqref{eq:ValleyC_Infinite} in contrast to the new definition that we will discuss next.
Topological quantities like $\bar{\nu}^P$
are only approximately quantized, unless the range of the integral in $k$ is infinite.
However, using $\Theta$
instead of $\bar{\nu}^P$ Eq. \eqref{eq:ValleyC_Infinite}
gives manifestly \emph{quantized} integers using a \emph{finite} loop about $P$.  
This property may be considered advantageous since we do not need an infinite area of 
integration.
Indeed, the relation to topology becomes clear when we  
show that the interband index is twice the valley Chern number, i.e., $\Theta=2 \bar{\nu}^P$
in 2-band $k\cdot p$ models.  

For illustration, consider the 2D gapped Dirac fermion model \cite{haldane1988model,zhang2018optical,zhang1919topological,dresselhaus2007group}, which 
has a $k \cdot p$ Hamiltonian with 
integer winding number $w$:
\begin{equation}
    \label{eq:hamiltonian:cf}
    \bm{H}(\bm{k}) = 
    \begin{pmatrix}
    \Delta & \alpha|\bm{k}|^\gamma e^{i w \phi_k}\\
    \alpha|\bm{k}|^\gamma e^{-i w \phi_k} & -\Delta 
    \end{pmatrix},
\end{equation}
where the energy gap is $2\Delta$, and
$\phi_{\bm{k}}=\tan^{-1}({k_y}/{k_x})$
\footnote{While $\gamma$ can take on arbitrary integral values in graphene
multilayers \cite{zhang2018optical}, 
we note that in
monolayer $\text{MoS}_2$ and gapped topological surface states,
$\alpha(|\bm{k}|)\propto|\bm{k}|=\sqrt{k_{x}^2+k_{y}^2}$
(that is, $\gamma=1$),
and that in biased bilayer graphene,
$\alpha(|\bm{k}|)\propto|\bm{k}|^2$ ($\gamma=2$).}.
The energy dispersions for the upper ($m$) and lower ($n$) bands are respectively
$\pm\sqrt{\Delta^2+\alpha^2|\bm{k}|^{2\gamma}}$.
For a circular loop parameterized as $ (k_x,k_y)=(r \cos(\lambda),r\sin(\lambda))$ and centered at the K' point (see Fig. \ref{fig:intro:interband} (c)),
the first three integrals in Eq. \eqref{eq:freq:onlyF}
conspire to give quantized $\Theta$.  
The last term in Eq. \eqref{eq:freq:onlyF}
does not exist in 2-band models.
As the area of integration  approaches the limit $k\rightarrow\infty$,
the third integral in Eq. \eqref{eq:freq:onlyF}
contributes less, making $\Theta$ the difference in integrals of 
${F}_m$ and ${F}_n$.
In the $k\rightarrow\infty$ limit, these two integrals are just $\bar{C}_m^{K}$ and $\bar{C}_n^{K}$. 
We demonstrate this in Fig. \ref{fig:intro:interband} (d), where we used $w=3$, $\Delta=1$, $\gamma=1$ and $\alpha=1$.
See SM \ref{app:CFberry} for more
on the model's Berry curvature.

To verify $\Theta= 2 \bar{\nu}^{K'}$, consider the $k\rightarrow\infty$ limit. The Berry curvature sum rule $\sum_i F_i=0$ gives $F_n=-F_m$ for 2-band models.
Since $\bar{\nu}^{K'} =\iint F_n$ in this limit, the claim follows from 
 Eq. \eqref{eq:freq:onlyF} 
 since the interband frequency integral tends to $0$.
Indeed, the figure shows $\iint F_n \rightarrow -1.5=\bar{\nu}^{K'}$. With $\Theta=-3$, we verify $\Theta= 2 \bar{\nu}^{K'}$.

\begin{figure}[h]
\centering
\valign{#\cr
  \hsize=0.495\columnwidth
  \begin{subfigure}{0.495\columnwidth}
  \centering
  \caption{}\label{fig:3a}
  \includegraphics[width=.95\textwidth,height=3.05cm]{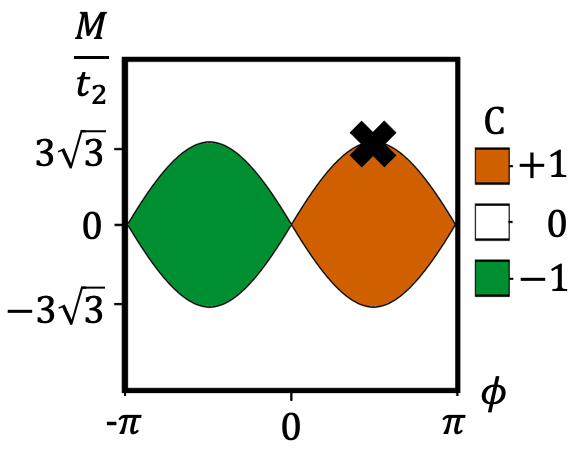}
  \end{subfigure}\vfill
  \vspace{1ex}
  \begin{subfigure}{0.495\columnwidth}
  \centering
  \caption{}\label{fig:3c}
  \includegraphics[width=.95\textwidth,height=3cm]{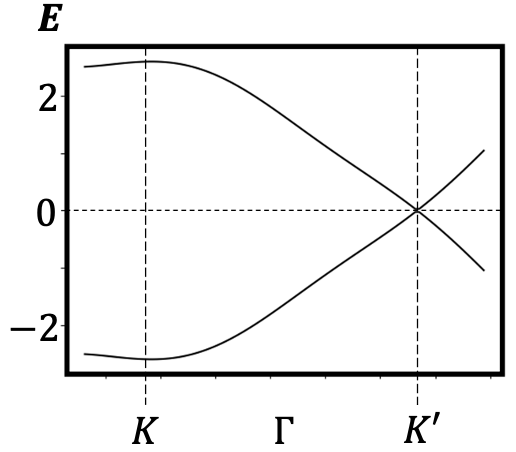}
  \end{subfigure}\cr
  \hsize=0.495\columnwidth
  \begin{subfigure}{0.495\columnwidth}
  \centering
  \caption{}\label{fig:3b}
  \includegraphics[width=.95\textwidth,height=3cm]{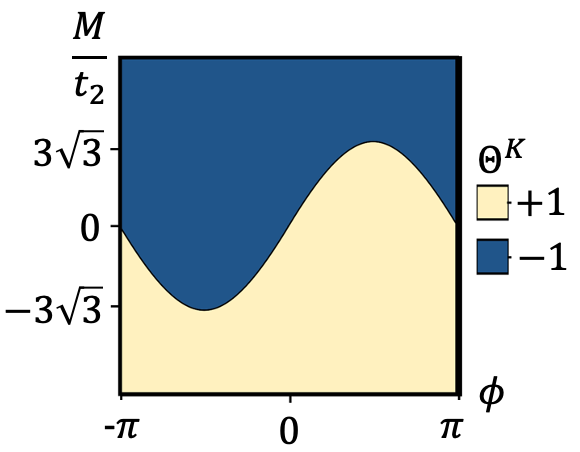}
  \end{subfigure}\vfill
  \vspace{1ex}
  \begin{subfigure}{0.495\columnwidth}
  \centering
  \caption{}\label{fig:3d}
  \includegraphics[width=.95\textwidth,height=3cm]{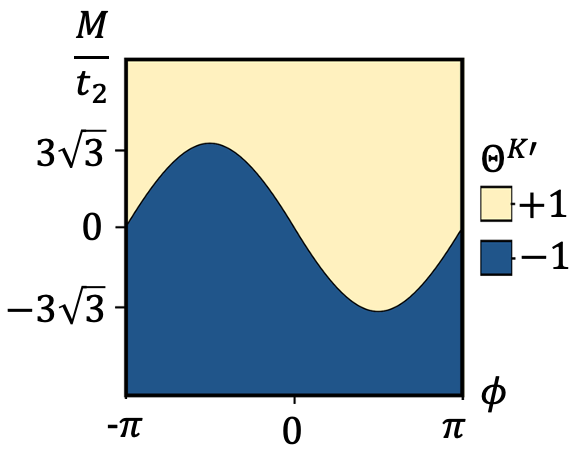}
  \end{subfigure}\cr
}
   \caption{    
        {(a)} A phase diagram of the first Chern number
        $C$ for the Haldane model as a function of $(M,\phi)$. 
        The topological phase transitions occur by gap closures at K or K'. 
        {(b)} Conduction and valence bands of the 
        Haldane model for $M/t_2=3 \sqrt{3}$, $\phi={\pi}/{2}$,
        and $t_1=4 t_2=1$. The gap closure at K' corresponds 
        to the
        phase boundary 
        marked with a cross $\bm{\times}$ in (a).
        {(c)} 
        $\Theta^K(M/t_2,\phi)$ using a fixed $k$-space loop 
        of radius $0.2$ around the K point 
        {(d)} $\Theta^{K'}(M/t_2,\phi)$.
        Notice that $\Theta^K(M/t_2,\phi)+\Theta^{K'}(M/t_2,\phi)=2 C
        (M/t_2,\phi)$, which is exactly twice 
        the expected phase diagram {(a)}.
        }
    \label{fig:HaldaneChern}
\end{figure}

Next, we discuss $\Theta$ in  
lattice models. 
For demonstration, we use 
Haldane's 2-band model for the 
quantum anomalous Hall effect \cite{haldane1988model}.
However, our results hold for all the models tested in 
SM \ref{app:topo_quantities}. 
On a honeycomb lattice,
its Hamiltonian can be written in a Bloch state basis on two sublattices $A,B$, using  Pauli matrices $\sigma_i$.
Below, 
$t_1$ is the nearest-neighbor hopping,
$t_2$ the amplitude of the complex second-neighbor hopping,
$\phi$ the phase accumulated by the $t_2$ hopping, and
$M$ the on-site energy between the $A$ and $B$ sublattices.
$\bm{a_i}$ are displacements from a $B$ site to its 
three nearest-neighbor $A$ sites, and
$\bm{b_i}$ are displacements for nearest-neighbor sites
in the same sublattices \cite{fruchart2013introduction}:
    \begin{eqnarray}
    \label{eq:Haldane:Hamiltonian}
        \bm{H}(\bm{k})
        &=&
        2 t_2 \cos\phi 
        \left[
        \sum_{i} \cos{(\bm{k}\cdot\bm{b_i})}
        \right] \mathbb \nonumber
        \\
        &+&
        t_1
        \left[
        \sum_{i}
        \left[
        \cos{(\bm{k}\cdot\bm{a_i})}\sigma_1
        +
        \sin{(\bm{k}\cdot\bm{a_i})}\sigma_2
        \right]
        \right] 
        \\
        &+&
        \left[
        M-2 t_2 \sin\phi
        \left(
        \sum_i \sin{(\bm{k}\cdot\bm{b_i})}
        \right)
        \right] \sigma_3. \nonumber
    \end{eqnarray}
This model can give 
topologically nontrivial first Chern numbers that may yield 
topologically-protected edge states \cite{haldane1988model}. 
The Chern number $C=\sum_{i\in\text{filled}}\quad\iint_{k\in \text{FBZ}} F_i$ (where $\text{FBZ}\equiv$ first Brillouin zone) changes when the band gap
closes and reopens at the high-symmetry points 
(K or K'),
as shown in Fig.  \ref{fig:HaldaneChern} (a)-(b). 
The physics at these valleys is therefore significant, because their gap closings can change the topology, and therefore edge state physics. 
However, unlike with the Dirac fermion model, it is not easy to define an analogous near-quantized topological quantity at valleys in lattice models. This is because the Berry curvature is not necessarily highly localized at $P$, and the area of the valley available for integration is finite. 
Therefore, quantities like $\bar{\nu}^{P}$ cannot often be directly
acquired from lattice models; at least not without low-energy expansions. 

However, $\Theta$ can again provide a quantized valley
characterization using a
small loop centered at $P$.
Figures \ref{fig:HaldaneChern} (c)-(d) show `phase diagrams'
analogous to Fig.  \ref{fig:HaldaneChern} (a), but
showing $\Theta$ for each valley. 
Clearly, when $\Theta^K$ and $\Theta^{K'}$ are
summed at each phase space $({M}/{t_2},\phi)$ point
in Fig. \ref{fig:HaldaneChern} (c) and (d),
we recover Haldane's phase diagram Fig. \ref{fig:HaldaneChern} (a): $\Theta^K+\Theta^{K'}=2 C$
\footnote{The factor of 2 is a peculiarity of 2-band models that
arises from the Berry curvature sum rule 
$\sum_{i\in\text{all bands}}
F_i=0$, which guarantees us that the Berry curvature at each
$k$-point of each band differs only by a sign. Therefore, 
the first two 
integrals in Eq. \eqref{eq:freq:onlyF} simplify to
a single integral with a factor of 2 (i.e., either $2\int_{{\mathcal{M}}}
	 {F}_{n} $ or $-2\int_{{\mathcal{M}}}
	 {F}_{m} $).}. 
Hence, compared to the state of the art, 
we now have a tool to analyze each valley in lattice models
without using low-energy expansions.

\bigskip

\emph{Band-resolved topological charge.}---The connections between the interband index 
and local topological characteristics
motivate us to define a band-resolved 
topological charge $\Xi^P$ for each $P$ that would add up to the Chern number $C$. 
For example, in the Haldane mode, we can define
$\Xi^K+\Xi^{K'}
\equiv{\Theta^{K}}/{2}+{\Theta^{K'}}/{2} = C$.
This band-resolved topological charge can be 
generalized to valley and multiband problems. 
This allows us to not only calculate band-resolved and valley topological indices, but also to
identify the number and source of edge states from inverting bulk bands without downfolding.

To make this multiband functionality apparent, 
we use $\Theta_{mn}$ (Eq. \eqref{eq:theta:k}) 
to define the novel generalized band-resolved topological charge $\Xi^\gamma_i$ per band $\ket{i}$, of 
\emph{context} $\gamma$ (e.g., $\gamma$ could be $P$, such as $K$).
For an $N$-level Hamiltonian, the
$N$ possible $\Xi_i^\gamma$ values may 
be found by solving the overdetermined simultaneous equations:
\begin{equation}\label{eq:simeqs}
    \{\Theta_{mn}^\gamma=\Xi_n^\gamma-\Xi_m^\gamma, \quad\sum_i\Xi_i^\gamma=0\}.
\end{equation}
We choose $N-1$ of the $N(N-1)/2$  equations involving $\Theta_{mn}^\gamma$
that make an appropriate linearly independent subset of equations 
along with 
$\sum_i\Xi_i^\gamma=0$, which is a 
\emph{conservation condition} 
that resolves linear dependence. 
This conservation condition is
analogous to the Berry curvature sum rule: 
for a complete basis, the topological charges should sum to $0$. 
Notice that since $\Theta_{mn}^\gamma$ is always an integer,
it is reasonable to expect
$\Xi_i^\gamma$ to be rational.

Motivated by how 
$\bar{\nu}^P$ is a sum over filled bands (Eq. 
\eqref{eq:ValleyC_Infinite}),
we define 
{\footnote{Recall that we used an overhead bar $\bar{\nu}^P$
in Eq. \eqref{eq:ValleyC_Infinite} to denote conventional
definitions. $\nu^\gamma$ in Eq. \eqref{eq:filled_bands_Xi}
lacks an overhead bar to differentiate our novel contribution, which is 
quantized as a rational number.}}: 
\begin{equation}
    \label{eq:filled_bands_Xi}
    \nu^\gamma=\sum_{i\in\text{filled}}\Xi_i^\gamma.
\end{equation}
We numerically find that $\Xi_i^{\gamma}$ and $\nu^\gamma$ are sufficient to calculate the number of edge modes, without using 
conventional topological quantities such as the Chern number and valley Chern number.
By the bulk-edge correspondence, 
the number of edge modes $b$
due to a domain boundary separating two 
systems $\alpha$ and $\beta$
is $b=|T^\alpha-T^\beta|$,
where $T^\gamma$ is
some topological character for context $\gamma$. For example, two adjacent 
Chern insulators 
give: $b=|C^\alpha-C^\beta|$
\cite{hasan2010colloquium}.
Or, for a domain boundary between two
valleys \footnote{{The domain boundary could be 
between 1D strips separating a $K$-edge from a $K'$-edge. Or
within one Brillouin zone, when inter-valley
scattering is suppressed \cite{xiao2010berry}.
}}, 
we have: $b=|\nu^K-\nu^{K'}|$
(Eq. \eqref{eq:filled_bands_Xi}) \cite{qiao2013topological,ezawa2014symmetry}.
And if the boundary is due to two bulk systems
with different external potentials $U_1$ and $U_2$ 
at the same $P$,
we get:
$b=|\nu^{P,U_1}-\nu^{P,U_2}|$ \cite{qiao2013topological}.
Our numerical results shown next support these claims. 
\\

We first exemplify the quantities we introduced using the gapped Dirac fermion model (Eq. \eqref{eq:hamiltonian:cf}).
For the example in Fig. \ref{fig:intro:interband} (d),
we use $\Theta=-3$ and solve the simultaneous equations $\{-3=\Xi_n-\Xi_m, \Xi_n+\Xi_m=0\}$ (Eq. \eqref{eq:simeqs})
to get $\Xi_n=-1.5$ and $\Xi_m=1.5$, which is consistent with 
the conventional valley Chern number per band (Eq.
\eqref{eq:ValleyC_Infinite}
${\bar{C}_n}^K\approx -1.5$
and ${\bar{C}_m}^K\approx 1.5$; see Fig. 
\ref{fig:intro:interband} (d)).

\begin{figure}[h]
    \begin{subfigure}{.49\columnwidth}
        \caption{}
        \includegraphics[width=0.85\linewidth]{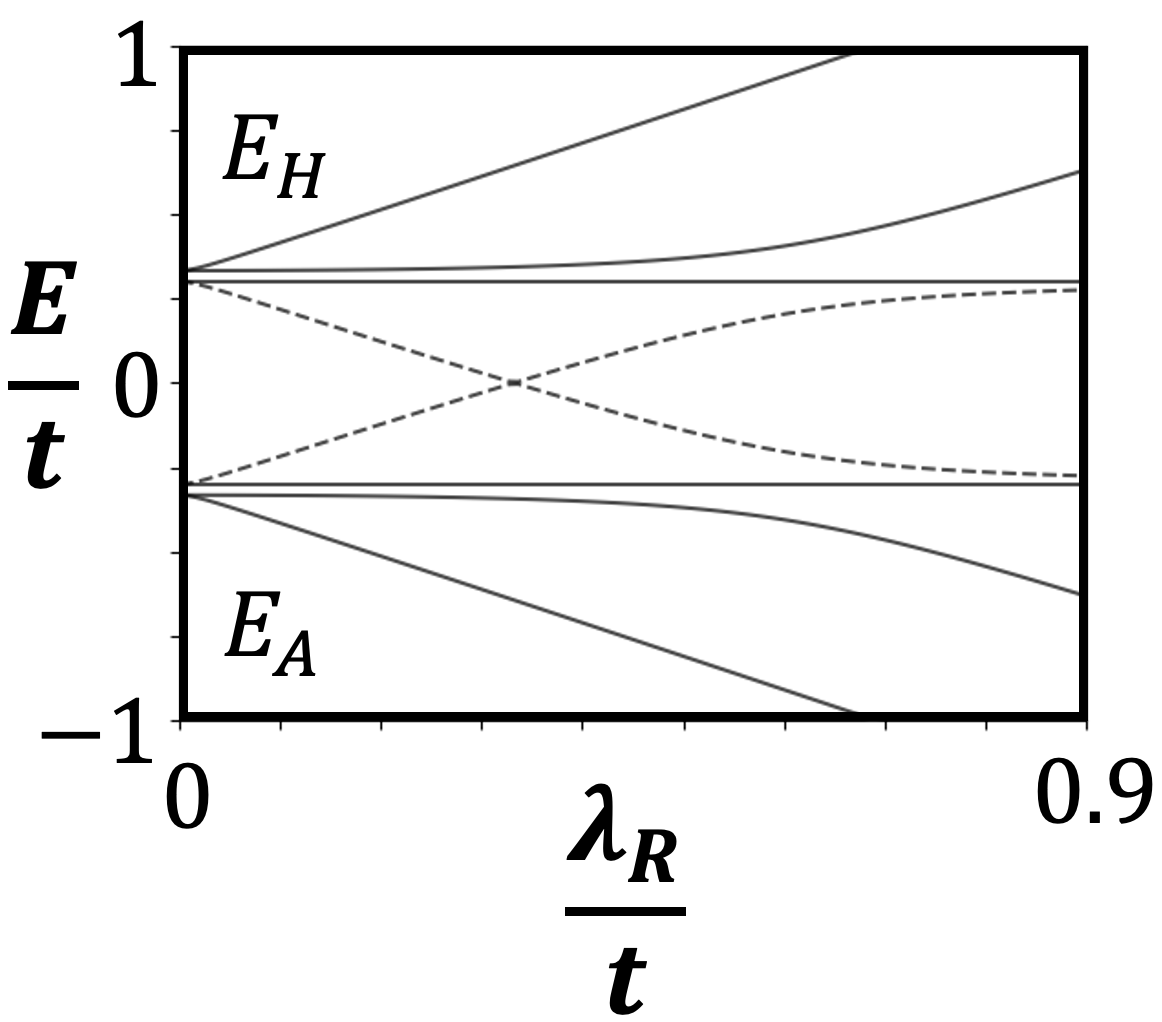} 
        \hfill
        \caption{}
        \includegraphics[width=\linewidth]{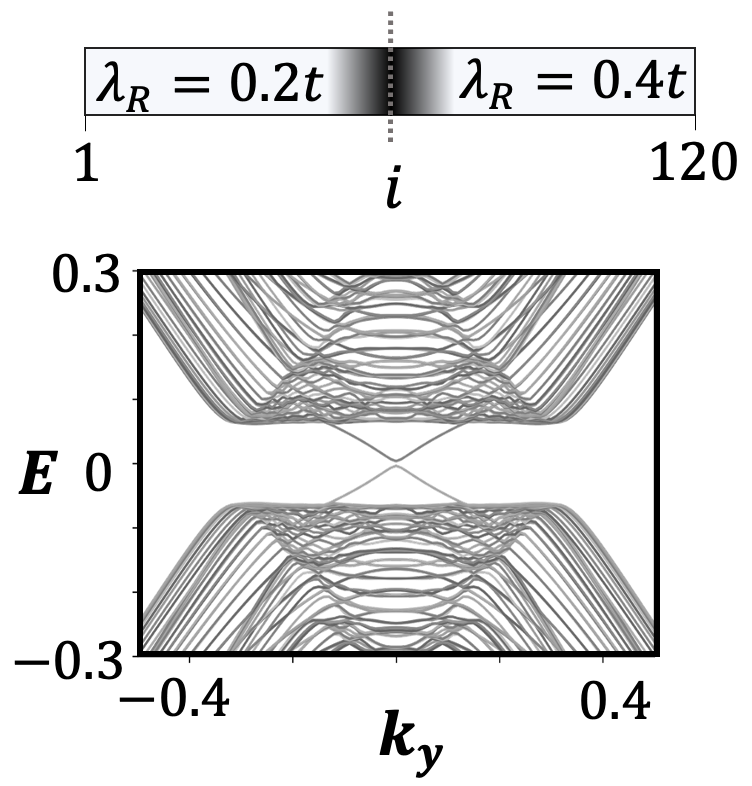}
        \hfill\hspace{10ex}
    \end{subfigure}
    \begin{subfigure}{.49\columnwidth} 
        \caption{}
        \includegraphics[width=\linewidth]{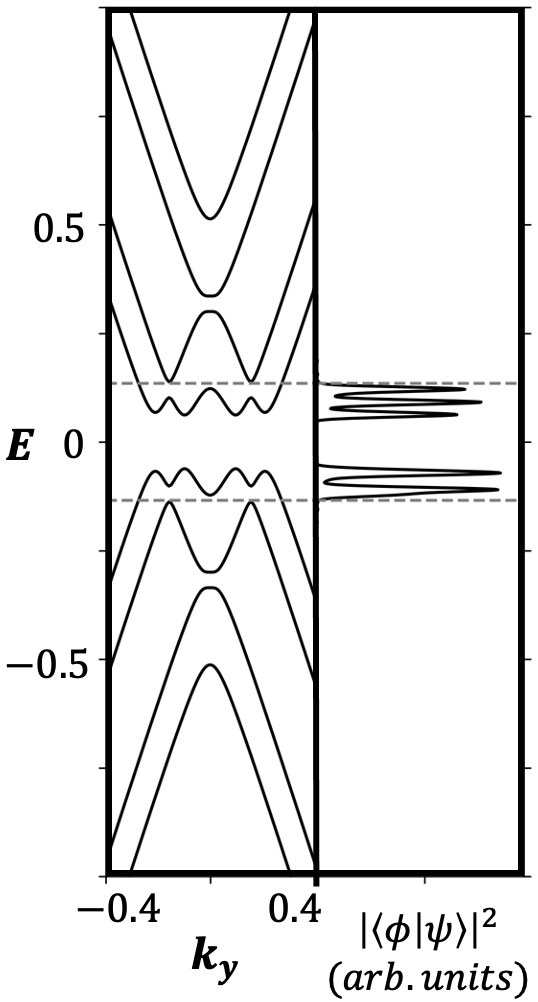}
        \vspace{4ex}
    \end{subfigure}
    \caption{
    {(a)} Band diagram at the $K$ point $(k_x,k_y)=(0,0)$ for
        the 8-band 
bilayer graphene model.
At $\lambda_R=0.2 t$, the bands are labeled $E_A, E_B, ..., E_H$ as we go from $-1$ to $+1$ along the vertical axis.
As $\lambda_R$ is varied from $0.2 t$ to $0.4 t$, the bands $E_D$ and $E_E$ invert at $\lambda_R\approx0.33 t$.
{(b)} \emph{Top:} Schematic of real-space nanoribbon with a domain boundary in the $x$-direction separating regions of two different $\lambda_R$. The shaded area represents the calculated wavefunction density. 
\emph{Bottom:} Nanoribbon bands at $k_x=0$ along the $k_y$ direction. 
{(c)}\emph{Left:} Bulk bands at $k_x=0$ for $\lambda_R=0.2 t$. \emph{Right:} Gaussian-broadened overlap element between bulk and domain boundary band wavefunctions
$|\braket{\phi | \psi}|^2$.
}
\label{fig:8bandstuff}
\end{figure}

For a multiband example, consider the 8-band model for
gated bilayer graphene including Rashba spin-orbit coupling 
(\cite{qiao2013topological}; parameter values in SM \ref{app:hamiltonians:8bandBLG}).
As the spin-orbit coupling parameter is tuned 
from $\lambda_R=0.2 t$ to $0.4 t$, 
we expect a band inversion at the $K$ valley \cite{qiao2013topological},
as in Fig. \ref{fig:8bandstuff} (a).
For these two values of $\lambda_R$,
we present $\Xi_i$
in Table \ref{tab:8-band}.

\setlength{\tabcolsep}{0.5em} 
{\renewcommand{\arraystretch}{1.5}
\begin{table}[]
\begin{tabular}{ccccccccc}
\hline\hline
\backslashbox{\,}{} & $\Xi_A$ & $\Xi_B$ & $\Xi_C$ & $\Xi_D$ & $\Xi_E$ & $\Xi_F$ & $\Xi_G$ & $\Xi_H$ \\ \hline
\multicolumn{1}{c}{$\Xi_i^{0.4}$} & $-\frac{1}{2}$ & $\frac{1}{2}$ & $-\frac{3}{2}$ & $\frac{1}{2}$ & $-\frac{1}{2}$ & $\frac{3}{2}$ & $-\frac{1}{2}$ & $\frac{1}{2}$ \\ 
\multicolumn{1}{c}{$\Xi_i^{0.2}$} & $-\frac{1}{2}$ & $\frac{1}{2}$ & $-\frac{3}{2}$ & $-\frac{1}{2}$ & $\frac{1}{2}$ & $\frac{3}{2}$ & $-\frac{1}{2}$ & $\frac{1}{2}$ \\ \hline\hline
\end{tabular}
\caption{
$\Xi_i^{\lambda_R}$ for the $K$ valley of the
8-band model for
gated bilayer graphene. At $K'$, $\Xi_i^{\lambda_R,K'}=-\Xi_i^{\lambda_R,K}$.
}
\label{tab:8-band}
\end{table}}

We map the $\Xi^{\lambda_R}_i$ in Table 
\ref{tab:8-band} to existing topological quantities: 
First, we see that for each choice of $\lambda_R$, $C=\nu^K+\nu^{K'}=0$, due to
$\Xi_i^{\lambda_R,K}=-\Xi_i^{\lambda_R,K'}$. 
This is consistent with the time-reversal symmetry of the model. 
We note that the limit of each 
$\bar{C}_i^{P}$ may not tend to the quantized
$\Xi_i$
in $N$-band models with $N>2$. To see this,
we calculated $\bar{C}_i^{P}$ at valley $K$ in the $k\rightarrow\infty$ limit using Eq. 
\eqref{eq:ValleyC_Infinite}.
For $\lambda_R=0.2 t$, $(\bar{C}^K_A,\bar{C}^K_B,\bar{C}^K_C,\bar{C}^K_D, ...) \approx (-0.01,0.99,-1.99,-0.99, ...)$.
We see that 
${\bar{C}_A}\approx -0.01 \rightarrow 0$, 
which is different from $\Xi_A=-1/2$.
This mismatch arises from the last two correction terms in Eq. \eqref{eq:freq:onlyF}, which do not 
necessarily tend to $0$ in models with $N>2$ bands. 

However, a difference between indices may indicate the number of edge states. Consider a valley problem with a domain boundary between $K$ and $K'$
for fixed $\lambda_R$ at half-filling.
From Table \ref{tab:8-band}, 
we have $b=|{\nu}^{K}-{\nu}^{K'}|=|-1-1|=2$ for $\lambda_R=0.4 t$,
and $b=|-2-2|=4$ for $\lambda_R=0.2 t$.
If we instead take a domain 
boundary problem at $K$ for the two $\lambda_R$ values,
we get $b=|\nu^{0.4}-\nu^{0.2}|
=|-1-(-2)|=1$ as the number of edge modes
due to the band inversion. These results are consistent with Ref. \cite{qiao2013topological}.

We also show that $\Xi_i$ can identify the bulk bands and orbitals causing edge states without explicitly tracking the evolution of spectra (as done in Fig. \ref{fig:8bandstuff} (a)). 
Since the only indices that differ between $\lambda_R=0.4 t$ and $\lambda_R=0.2 t$ are $\Xi_D$ and $\Xi_E$, the orbitals causing edge states are from bands $E_D$ and $E_E$. 
To verify this, we further model a tight-binding nanoribbon 
of 120 sites that is periodic in the $y$-direction, that has a domain boundary in the $x$-direction at site $i=60$, as in Fig. \ref{fig:8bandstuff} (b) (top).
We discretized the continuum model 
Eq. \eqref{eq:8bandHamiltonian} to get a tight-binding model that includes both valleys. 
Our calculations \cite{groth2014kwant} show that edge states accumulate at the domain boundary.
We then calculated the nanoribbon bands in Fig. \ref{fig:8bandstuff} (b) (bottom), which shows one zero-energy edge state from each valley. 
Fig. \ref{fig:8bandstuff} (c) shows that these edge states are composed of
orbitals from $E_D$ and $E_E$, evident from
the large overlap between the 
zero-energy edge state wavefunction $\ket{\phi}$
and the bulk wavefunctions $\ket{\psi}$ of $E_D$ and $E_E$ (calculated in a homogeneous nanoribbon at $\lambda_R=0.2 t$). 
\\

\emph{Conclusion and outlook.}---We introduced two gauge-invariant quantities, the interband index $\Theta_{mn}$ and band-resolved topological charge $\Xi_i$. These quantized indices offer novel characterizations of topologically significant submanifolds in 2D $k$-space manifolds that are
consistent with existing topological characters such as the first and valley Chern numbers. 
As demonstrated, the differences between $\Xi$ values from different contexts may carry desired physical meaning
as the number of edge states. So,
the universality and significance of individual $\Xi$ warrant further investigation, as does the interpretation of $\Theta$ and $\Xi$ for loops not enclosing a single $P$
(see SM \ref{app:kloops}). The non-Abelian version of the interlevel index provided in Ref. \cite{xu2018quantized} may be extended to treat degeneracies, and deserves further work due to the prevalence of accidental and symmetry-protected degeneracies in condensed matter systems.
In conclusion, these first-in-literature quantities, due to their elegant, quantized nature and broad applicability, are prime candidates for deeper study.

We acknowledge helpful discussions with Chao Xu from UC San Diego, 
and with Yafei Ren and Di Xiao from the University of Washington.

\bibliography{IB2_references}

\newpage
\section*{Supplementary Material}

\appendix

\section{\label{app:expanddarg} Expanding the $\dif \arg$ term: deriving interband frequency}

In this section, we explore $\Theta$ analytically using the mathematical relation 
$\dif \arg z= \text{Im}\frac{\dif z(\lambda)}{z(\lambda)}$ 
for $z(\lambda) \in \mathbb{C}\setminus\{0\}$.
We now define the overhead dot $\dot{ }$ and $\dif$ to both
denote derivation with respect to $\lambda$.
To prove this relation, notice that $z=|z|e^{i \textrm{Arg}(z)}$ 
implies that $\log z = \log |z| + i \textrm{Arg}(z)$.
Although $\textrm{Arg}$ is defined only up to a constant multiple of $2\pi$ ($\equiv c$), 
this ambiguity disappears when we take the
derivative: $\frac{\dif}{\dif \lambda}\textrm{arg}(z)=\frac{\dif}{\dif \lambda}(\text{Im} \log \frac{z}{|z|} + c)=\text{Im}\frac{\dot{z}}{z}$.
Note that the derivative of $\textrm{arg}(z)$ is well-defined locally when 
$z \in \mathbb{C}\setminus\{0\}$.
Consider a general $N$-level model with non-degenerate bands.
Then, for $z(\lambda)=\langle m | \dot{n} \rangle$:
\begin{align}
\label{eq:applydargmath}
\begin{split}
&\dif \arg \langle m|\dot{n}\rangle = \text{Im}\frac{\dif \langle m|\dot{n}\rangle}{\langle m|\dot{n}\rangle},
\\
&\text{and }\text{Im}(\dif \langle m|\dot{n}\rangle) = \text{Im}(\langle \dot{m}|\dot{n}\rangle + \langle m|\ddot{n}\rangle).
\end{split}
\end{align}
Then, for levels $\ket{m},\ket{n}$ and $\ket{q}$,
\begin{align*}
\langle \dot{m}&|\dot{n}\rangle = \sum\limits_{q} \langle \dot{m}|q\rangle\langle q| \dot{n}\rangle \\
&=
\langle \dot{m}|m\rangle\langle m| \dot{n}\rangle + \langle \dot{m}|n\rangle\langle n| \dot{n}\rangle 
	+ \sum\limits_{q\neq m,n}\langle \dot{m}|q\rangle\langle q| \dot{n}\rangle \\
&= - \langle m|\dot{m}\rangle\langle m| \dot{n}\rangle 
- \langle m|\dot{n}\rangle\langle n| \dot{n}\rangle
- \sum\limits_{q\neq m,n}\langle m|\dot{q}\rangle\langle q| \dot{n}\rangle \\
&= \langle m| \dot{n}\rangle \left[i\mathcal{A}_m + i\mathcal{A}_n 
- \sum\limits_{q\neq m,n}\frac{\langle m| \dot{q}\rangle\langle q| \dot{n}\rangle}{\langle m| \dot{n}\rangle}\right],
\end{align*}
where we used the definition for the Berry connection $\langle a | \dot{a}\rangle = -i \mathcal{A}_a$ and the
relation $ \dif\langle a | b \rangle = \dif (0)$ implies that $\langle \dot{a} | b \rangle = -\langle a | \dot{b} \rangle $
for orthonormal states $\{\ket{a},\ket{b}\}$.

To treat the last term in \eqref{eq:applydargmath},
recall that for any two functions $f,g$ of $\lambda$, the second derivative is
$\dif^2(f g)=(\dif^2 f)g+ 2 (\dif f) (\dif g) + f(\dif^2 g)$. Applying this
to Schr\"{o}dinger's equation 
$H|n\rangle=E_n |n\rangle$ gives:
$$
\ddot{E}_n|n\rangle+ 2 \dot{E}_n |\dot{n}\rangle + E_n | \ddot{n}\rangle \\
= \ddot{H}|n\rangle+ 2 \dot{H}|\dot{n}\rangle + H | \ddot{n}\rangle.
$$
Applying ket $\langle m|$, using $\langle m | n\rangle = 0$ and the notation $E_{nm}=E_n-E_m$:
$$
\langle m | \ddot{n} \rangle = \frac{\langle m|\dot{n}\rangle}{E_{nm}}\left(-2 \dot{E}_n 
+ \frac{\langle m |\ddot{H} | n\rangle}{\langle m|\dot{n}\rangle} + 2\frac{\langle m |\dot{H}|\dot{n}\rangle}{\langle m|\dot{n}\rangle}\right).
$$
For a general $N$-level Hamiltonian, the numerator of 
the last term above becomes:
\begin{align*}
&\langle m |\dot{H}|\dot{n}\rangle = \sum\limits_{q} \langle m |\dot{H}|q\rangle\langle q | \dot{n}\rangle \\
&=
\langle m |\dot{H}|m\rangle\langle m |\dot{n}\rangle
+
\langle m |\dot{H}|n\rangle\langle n |\dot{n}\rangle
+
\sum\limits_{q\neq m,n}\langle m |\dot{H}|q\rangle\langle q |\dot{n}\rangle.
\end{align*}
Using the Hellman-Feynman-type relation $\langle m |\dot{H}|n\rangle = E_{nm} \langle m |\dot{n}\rangle$, the above
$$
= \langle m|\dot{n}\rangle \left(\langle m |\dot{H}|m\rangle + E_{nm}\langle n | \dot{n}\rangle 
+ \sum\limits_{q\neq m,n}E_{qm}\frac{\langle m|\dot{q}\rangle\langle q|\dot{n}\rangle}{\langle m |\dot{n}\rangle}\right).
$$
For $\beta=\{\langle m |\dot{H}|m\rangle\}$ with $\dot{E}_n \in \mathbb{R}$, we have $\text{Im}(\beta)=0$. 
Finally, we get:
\begin{align*}
&\text{Im}\frac{\dif \langle m|\dot{n}\rangle}{\langle m|\dot{n}\rangle}\\
&=
\mathcal{A}_m - \mathcal{A}_n + \text{Im}\frac{\langle m |\ddot{H} | n\rangle}{\langle m|\dot{H}|n\rangle}\\
&\quad+
\sum\limits_{q\neq m,n}\text{Im}\frac{\langle m|\dot{q}\rangle\langle q|\dot{n}\rangle}{\langle m |\dot{n}\rangle}\left(2 \frac{E_{qm}}{E_{nm}} - 1\right) \\
&= 
\mathcal{A}_m - \mathcal{A}_n + \text{Im}\frac{\langle m |\ddot{H} | n\rangle}{\langle m|\dot{H}|n\rangle}\\
&\quad+
\sum\limits_{q\neq m,n}\text{Im}\frac{\langle m|\dot{H}|q\rangle\langle q|\dot{H}|n\rangle}{\langle m |\dot{H}|n\rangle}\left(\frac{2}{E_{nq}} - \frac{E_{nm}}{E_{qm}E_{nq}}\right).
\end{align*}
When plugged into Eq. \eqref{eq:theta:k}, we get Eq. \eqref{eq:freq:onlyF} given in the main text.

\section{\label{app:CFberry}Berry curvature of gapped Dirac model}
\begin{figure}[h]
    \includegraphics[width=0.5\textwidth]{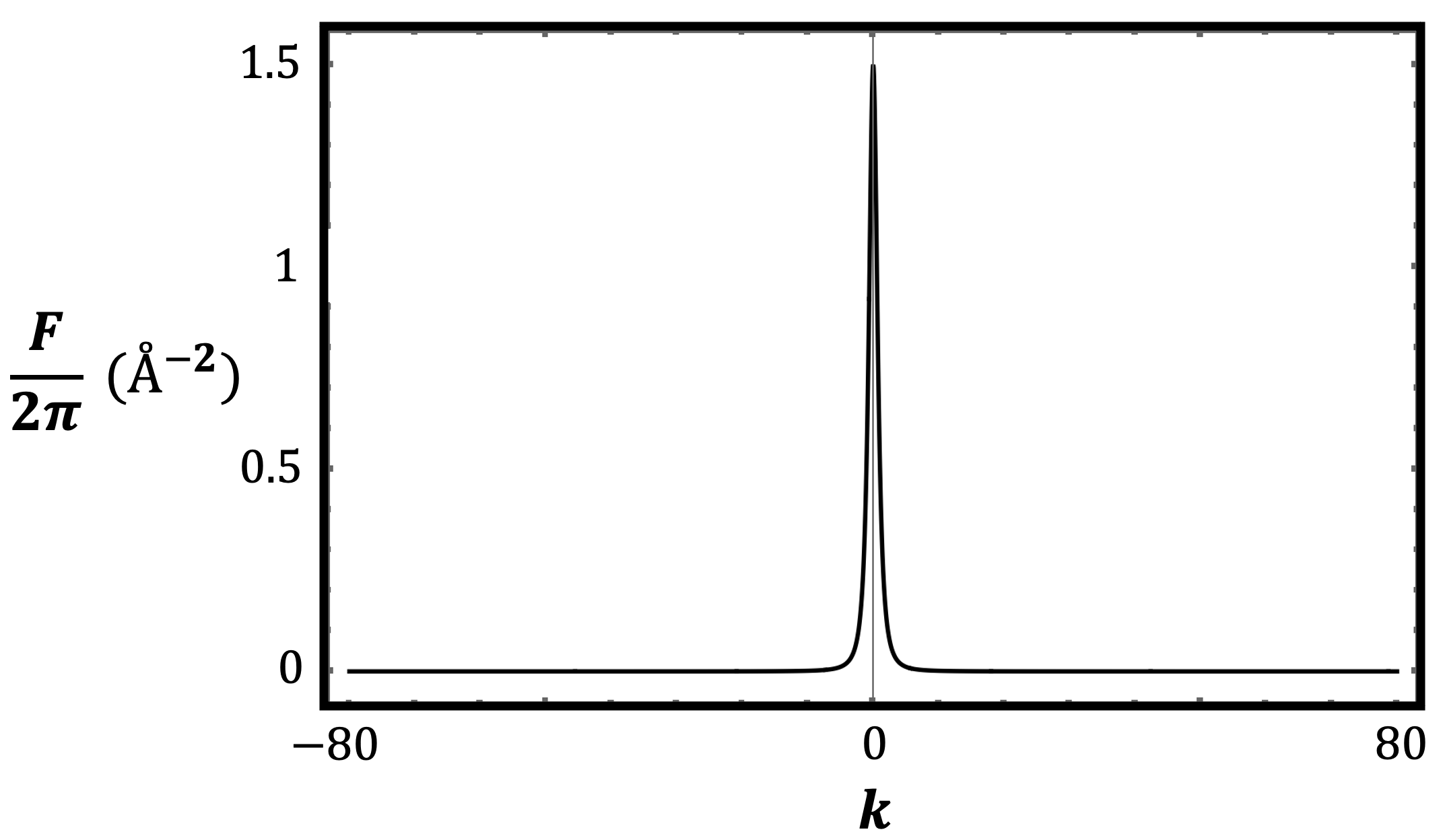}
    \caption{      
    Berry curvature of the gapped Dirac fermion model 
    along the line $k_y=0$. This is the same case as in 
    Fig.  \ref{fig:intro:interband} (d), where varying $r$
    from 0 to 80 encompasses $-80\leq k_x \leq 80$. 
    }
    \label{fig:CFberry}
\end{figure}

Using Eq. \eqref{eq:hamiltonian:cf}, the Berry curvature of the 
gapped Dirac fermion model is calculated as \cite{xiao2010berry}:
\begin{equation}
	\label{eq:app:CF:berryC}
    \frac{\gamma \omega \Delta \alpha^2 |k|^{2(\gamma-1)}}
    {2(\Delta^2+\alpha^2 |k|^{2\gamma})^{\sfrac{3}{2}}}.
\end{equation}
Figure \ref{fig:CFberry} shows the Berry curvature distribution to
supplement the argument around Fig. \ref{fig:intro:interband} (d). Notice that the Berry curvature
sums to $\approx 1.5 \times 2\pi$ \AA$^{-2}$.

\section{\label{app:topo_quantities}Models used; and definitions of first, valley, spin and spin-valley Chern numbers}

We tested the following $k\cdot p$ models:
2-band gapped Dirac fermion model \cite{zhang2018optical};
4-band gapped bilayer graphene with layer-stacking wall \cite{zhang2013valley};
4-band spinful Dirac Hamiltonian on monolayer honeycomb lattice \cite{ezawa2014symmetry}; 4-band ferrimagnetic honeycomb lattice model for the valley-polarized quantum anomalous Hall effect \cite{zhou2017valley22}
(see also \cite{pan2014valley,vila2021valley,zhou2021quantum});
6-band ABC-Stacked trilayer graphene \cite{koshino2009trigonal};
8-band gated bilayer graphene with Rashba spin-orbit coupling and exchange field (\cite{qiao2013topological}, \ref{app:hamiltonians:8bandBLG}). 
We tested the following lattice models: 
Haldane model for the quantum anomalous Hall effect \cite{haldane1988model};
the
3-band model in \cite{liu2018generalized};
and models allowing $|C|>1$, even if 
they have more than 2 
high-symmetry points $P$ beyond K and K'
(such as in \cite{sticlet2012geometrical}). 

We note that electron-hole symmetry is not necessary for our conclusions to hold. We verified this using the 3-band model \cite{liu2018generalized}, which inherently lacks electron-hole symmetry. 

To define existing topological indices
using a specific example, consider a spinful system
with only two $P$ ($K,K'$) for each 
spin label ($\uparrow,\downarrow$). This could be
the Dirac Hamiltonian on a monolayer honeycomb lattice \cite{ezawa2014symmetry}. 
Then, per \cite{ezawa2014symmetry},
the first, valley, spin, and spin-valley Chern numbers are respectively:
\begin{align}
    \begin{split} \label{eqs:chernDEFINITIONS}
    {C} &= \nu^{K,\uparrow} + \nu^{K',\uparrow} + \nu^{K,\downarrow} + \nu^{K',\downarrow}\\
    V &= (\nu^{K,\uparrow} + \nu^{K,\downarrow}) - (\nu^{K',\uparrow} + \nu^{K',\downarrow})\\
    2 {C}_s &= (\nu^{K,\uparrow} + \nu^{K',\uparrow}) - (\nu^{K,\downarrow} + \nu^{K',\downarrow})\\
    2 {C}_{sv} &= (\nu^{K,\uparrow}+ \nu^{K',\downarrow})- (\nu^{K',\uparrow} + \nu^{K,\downarrow}).
    \end{split}
\end{align}
Under certain symmetries, the $\mathbb{Z}_2$ index 
for topological insulators $={C}_s$ (mod 2) 
\cite{ezawa2014symmetry}.
For all models tested
{\footnote{With the exception of \cite{liu2013three},
which remains an open problem.}},
we recovered the expected topological 
quantities \eqref{eqs:chernDEFINITIONS}.
For instance, we got the $N\pi$ Berry
phases for the $N$-layer graphene models, 
and the $V$ in \cite{zhang2013valley}.
Results not included in this paper
may be requested from the authors upon reasonable request.

\section{\label{app:TmnEQMTnm}Showing $\Theta^{mn}=-\Theta^{nm}$}

First we consider $\Delta\Phi$ in \eqref{eq:theta:k}:
$
\Delta\Phi^{mn} = \Phi_m - \Phi_n = -(\Phi_n - \Phi_m) = -\Delta\Phi^{nm}.
$
Next, we consider the term $\braket{m|\dot{n}}$ (using the $\lambda$ formulation for simplicity). Notice that:
$
\dif (0) = \dif(\braket{m|n}) = \braket{\dot{m}|n}+\braket{m|\dot{n}}=0.
$
It follows that $\braket{\dot{m}|n}=-\braket{m|\dot{n}}$.
Putting all this together, we get $\Theta^{mn}=-\Theta^{nm}$.

\section{\label{app:kloops}Choosing $k$-space loops}

In the main text, we 
considered $\Xi_n^P$ per high-symmetry point $P$. This means that 
loops used to calculate 
$\Xi$ must enclose only one special point $P$.
Based on our numerical results, these $P$ 
are high-symmetry points associated with crystal 
symmetry
(such as $K,K',M,\Gamma,X,...$). 
They may not necessarily be
points where the band gap may close.
We observe that the interband matrix element
$\Braket{{m} | \boldsymbol{\nabla_k} n}\cdot
	\hat{e}_{\tau}$ \eqref{eq:theta:k}
diverges at these $P$. However, 
due to the matrix element's gauge-dependence,
we doubt that this observation can be used to 
identify $P$. This is also due to 
the fact that, in a non-azimuthal basis, 
we do not have a unique $k$-space vector 
field prior to choosing a loop. Otherwise,
its critical points may have been used to identify $P$.
We get the results in this work 
only for loops enclosing high-symmetry points. 
Why these points $P$ are important 
remains an open question.

Another numerical observation is that 
the winding loop of the interband matrix element
is coincidentally discontinuous if the $k$-space loop
does not include only Berry curvature $F$ of only one sign. Our numerical implementation demonstrated this
discontinuity also when the band is gapless
anywhere in parameter space (not necessarily inside 
or along the loop). This is very likely a numerical 
artefact.
Our calculations also indicate that loops 
not enclosing aforementioned high-symmetry points 
$P$ may only yield $\Theta=\pm 1$.
In 2-band models, we see that
$\Theta\propto\text{sign}(F \text{ of the region 
within the loop})$.

\section{\label{app:hamiltonians:8bandBLG}Gated bilayer graphene with Rashba spin-orbit coupling and exchange field}

Ref. \cite{qiao2013topological}
gives the following momentum space 
8-band effective Hamiltonian for the valley points
$K$ and $K'$ (centered at the origin $(0,0)$ and respectively
labeled by $\eta=\pm 1$):
\begin{align}
\label{eq:8bandHamiltonian}
\begin{split}
\bm{H}(\bm{k}) 
&= v(\eta\sigma_z k_x + \sigma_y k_y)\bm{1}_s
\bm{1}_\tau
+ \frac{t_\perp}{2}(\sigma_x\tau_x-\sigma_y\tau_y)\bm{1}_s \\
&+ \frac{\lambda_R}{2}(\eta\sigma_x s_y - \sigma_y s_x)\bm{1}_\tau
+ M s_z \bm{1}_\sigma\bm{1}_\tau + U \tau_z
\bm{1}_s \bm{1}_\tau,
\end{split}
\end{align}
 where $s$, $\sigma$ and $\tau$ are Pauli matrices 
representing the spin, AB sublattice, and top-bottom layer degrees of freedom respectively. $ \bm{1}_i$ are $2\times 2$ identity matrices.
The Fermi velocity is given by 
$v=3 a t / 2$ with $a$ the lattice constant, and $t$ the hopping amplitude.
$t_\perp$ is the interlayer tunneling amplitude, $\lambda_R$ the Rashba spin-orbit coupling, $U$ the interlayer potential, and $M$ the exchange field.
We used $a=1, t=2.6, t_\perp = 0.143 t, M=0,U = 0.3 t$ and $t_R = 0.058 t$. 

We ignored the exchange field $M$ in our calculations for simplicity in presentation. However, using $M\neq 0$, we were able to reproduce the phase diagrams in Ref. \cite{qiao2013topological} which classify quantum anomalous Hall, quantum valley Hall, and metallic phases (c.f. Figure 15 of the reference). 

For our numerical calculations, we used \emph{MATLAB} and \emph{Python}. We
used $k$-loops discretized to $250$ segments, and
a $100\times100$ $k$-grid.


\end{document}